# Electromagnetic fields created by the lightning leader channel in the final jump phase

N.I. Petrov and G.N. Petrova

The dependences between the parameters of the Lightning discharge in the leader phase and the return stroke phase are obtained. It is shown that the peak values of the electric field created by lightning are determined by the parameters of the streamer zone of the leader at the final jump phase. It is shown that a large peak value and the steepness of the lightning return stroke current are generated in the final jump phase.

## 1. INTRODUCTION

Lightning radiation has a maximum intensity in the area of 5-20 $kHz$ and its spectral density is inversely proportional to the frequency [1]. The wavelength corresponding to the maximum intensity of radiation is of the order of the length of Lightning channel. The second characteristic size of Lightning channel is the characteristic length of channel bending, which corresponds to the frequency range of $f = 3 \div 300$ $MHz$ (wavelength $\lambda = 1 \div 100$ $m$).

There is another characteristic size – the cross size of Lightning channel, corresponding to the frequency range $f = 300 \div 3000$ $MHz$ ($\lambda = 0.1 \div 1$ $m$). There is also the random bending of leader channel in this range.

If the frequency range $5 \div 100$ $kHz$ may be well described as a radiation of linear dipole, then the higher frequency radiation cannot be considered in a dipole approximation.

In this paper the functional dependences between different parameters of Lightning and its electromagnetic field are obtained.

## 2. ELECTROMAGNETIC FIELD GENERATED BY LIGHTNING DISCHARGE

The characteristics of lightning radiation are closely related to the spatio-temporal changes of charges and currents. Therefore, to calculate the electromagnetic field of lightning, it is necessary to know the parameters of the current pulses flowing during discharge. As a rule, lightning is represented as a vertical channel through which a current pulse moves. The vertical electric field intensity at ground level ($z = 0$) produced by the vertical channel above the perfectly conducting surface shown in Fig. 1 in a dipole approximation in a far field zone is determined by the expression [1, 2]:

$$E(x,y,0,t) = \frac{1}{2\pi\varepsilon_0} \left[ \begin{array}{l} \int_0^H \frac{(2-3\sin^2\theta)}{R^3} \int_0^t i(z,\tau - R/v)d\tau dz + \int_0^H \frac{(2-3\sin^2\theta)}{cR^2} i(z,t-R/v)dz \\ - \int_0^H \frac{\sin^2\theta}{c^2 R} \frac{\partial i(z,t-R/v)}{\partial t} dz \end{array} \right] \quad (1)$$

where $\varepsilon_0$ is the permittivity of free space, $i(z, t)$ is the channel current, $H$ is the radiating length of the channel, $R$ is the distance up to the observation point, $v$ is the speed of the pulse current, $c$ is the speed of light and the angle is defined in Fig. 1.

The expression (1) is valid, when $R \gg H$. The first term in (1) corresponds to the electrostatic field, the second – induction field, and the third – radiation field. The radiation term is significant at large $R$, when other terms can be neglected.

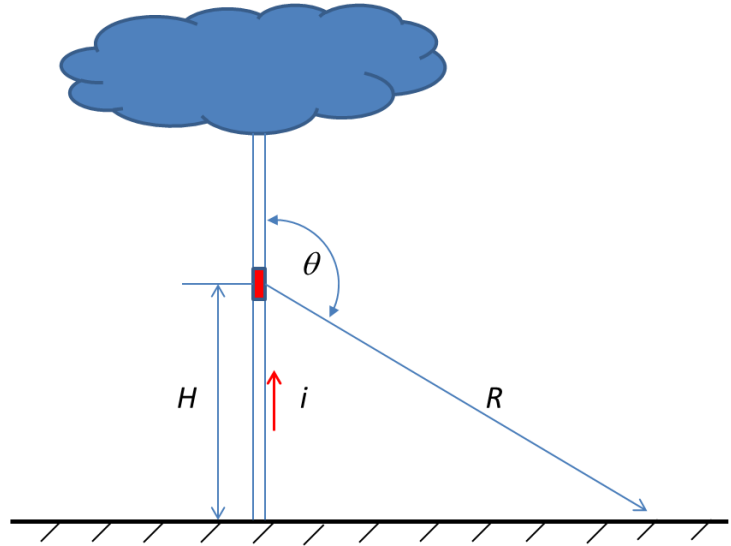

Fig. 1. Vertical lightning channel above a flat perfectly conducting surface.

Thus, it is necessary to know the space-time distribution of charges and currents in lightning discharge in order to calculate the potentials and electric fields. However the difficulties of measurements of electric and magnetic field intensities directly in the area of discharge did not allow determination the spatial distribution of charges along the lightning channel yet. Therefore the measurements of electric field created by leader discharge in a laboratory are important for receiving information about the structure of leader channel and construction of physical model of lightning. Results of such measurements have been presented in [3 - 5].

The change in the intensity of the electrostatic field in the discharge gap is due to the time dependence of the electrode potential and the dynamics of the spatial charges of the streamer zone and the leader channel.

The waveform of the electric field measured using Pockel's device in the rod-plane discharge gap with length $H = 6$ m at a positive polarity of the applied voltage is presented in Fig. 2.

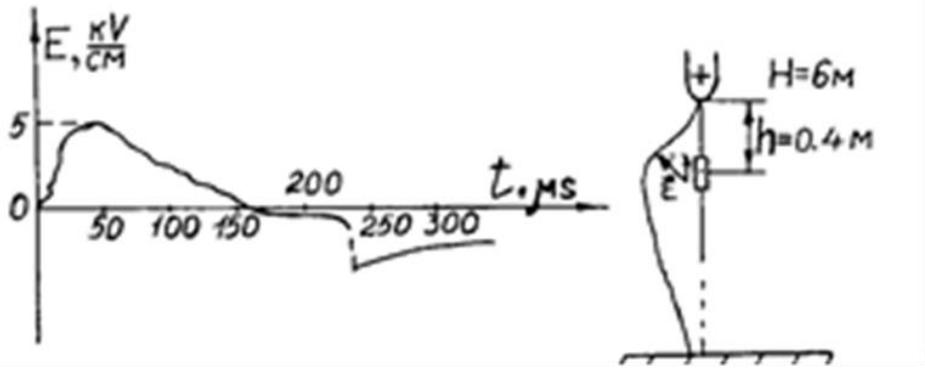

Fig. 2. Measured electrostatic field waveform nearby the positive leader channel in air gap [5].

One can distinguish two stages in waveforms, which correspond to different physical processes. First stage corresponds to the leader phase of a discharge, second stage - to the neutralization of the space charges introduced into the gap (return stroke phase). Between this two stages breakdown (return stroke) occurs when the amplitude of the field increases sharply up to the value 5 kV/cm during the time less than 1 µs. Electric field decreases smoothly after breakdown. The duration of the leader phase is determined by the leader propagation velocity and it increases with the growth of a gap length.

For negative polarity of the applied voltage the electric field intensity has negative polarity at the leader phase and positive polarity at the return stroke phase (Fig. 3).

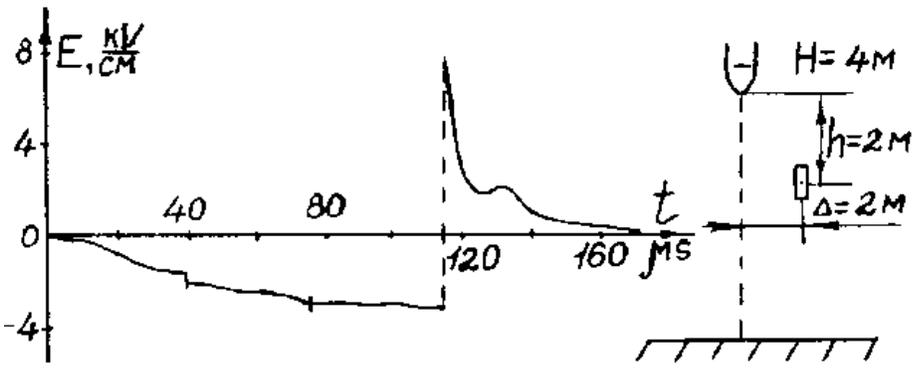

Fig. 3. Measured electrostatic field waveform nearby the negative leader channel in air gap with length $H = 4$ m [5].

It is seen that there are sharp changes in the electric field caused by the step propagation of the negative leader. The time intervals between these rapid changes in the electric field correspond to the time of the step formation and it is about 40 μs. Note that the neutralization time of spatial charges is significantly less than for positive polarity discharges. This is also the case for lightning discharges.

As follows from the waveforms, the submicrosecond changes of the field are caused both by the formation of steps in a leader phase and formation of the final stage of breakdown.

In Fig. 4 the streak photograph of a positive leader/streamer development in a rod-plane gap is presented. The gap length $H = 10$ m, the leader developing time up to breakdown $T_{br} = 410$ μs.

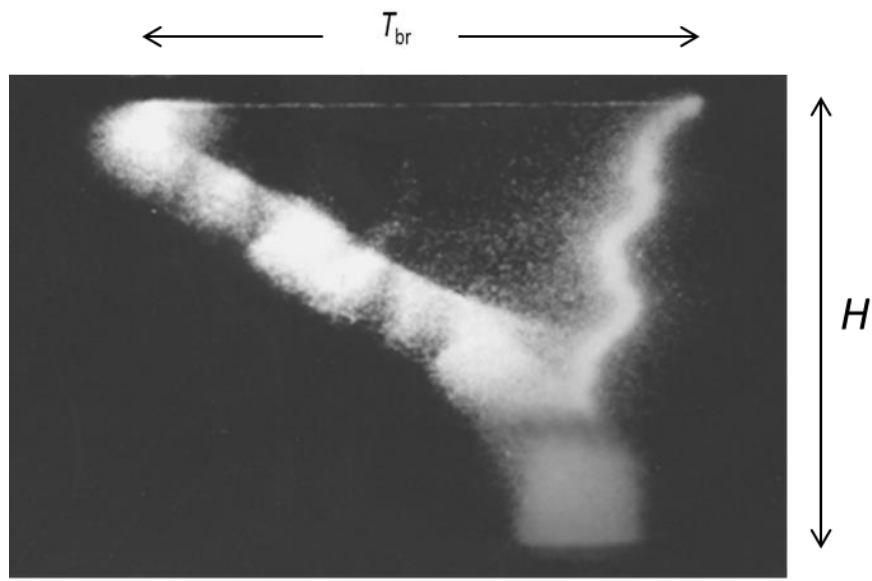

Fig. 4. Streak photograph of a positive leader/streamer developing in a long rod-plane gap [5].

Submicrosecond field variations are due to the fast changes of currents caused by the dynamics of space charges. These high speed changes in the current take place both at the leader phase and the return stroke phase of lightning discharge. At the leader phase these changes are connected with the formation of steps (Fig.3).

According to Eq.1 rapid changes in current in the submicrosecond range cause a field of high-frequency (HF) radiation in the far field region. This indicates that the physical mechanism of HF radiation from lightning discharges may be associated with the step propagation of the leader and its final jump phase.

## 3. EVALUATION OF LEADER CURRENT

The peak value and the steepness of the lightning discharge current are generated at the return stroke phase. In its turn, the formation of the return stroke current front is determined by the final jump phase of a leader.

The current flowing in the leader channel until it touches the ground surface can be represented through the displacement current at the front of the streamer zone [6]:

$$I \approx \varepsilon_0 \frac{\partial E}{\partial t} S \approx \varepsilon_0 \frac{\partial E}{\partial x} v_{sf} S \approx \pi^2 \varepsilon_0 E_s l_s v_{sf}, \qquad (2)$$

where $S$ is the surface of the streamer zone front, $l_s$ is the streamer zone length, $v_{sf}$ is the velocity of streamer zone front, $E_s$ is the electric field intensity at the streamer zone front, $\varepsilon_0 = 8.85 \cdot 10^{-12}$ F/m is the dielectric constant.

The velocities of a streamer zone front and leader in the leader propagation phase are the same. For the leader velocity $v_l \approx 2 \div 20$ cm/µs and streamer zone length $l_{tsr} \approx 3 \div 10$ m we have the leader current

$I_l \approx 3 \div 100$ A.

At the final jump phase the streamer zone length and the velocity of the streamer zone front are sharply increased, that causes the growth of the current.

Computer modeling of propagation of the leader channel – streamer zone system taking into account the influence of the "image" charges shows a sharp lengthening of the streamer zone at the final jump phase [7]. The streamer zone length increases 2-3 times as soon as a few microseconds. Note that the calculation results agree well with the laboratory experiments.

The lengths of the streamer zone at the final jump phase may become a few ten up to a few hundred meters and the velocity of the streamer zone front reaches the values of $10 \div 100$ m/µs. Substituting these values into (2) we obtain that the currents caused by the final jump phase equal to $I_0 \approx 5 \div 500$ kA.

For the current derivative d$I$/d$t$ we have

$$\dot{I} \approx \pi^2 \varepsilon_0 E_s \left( v_{sf}^2 + l_s \frac{dv_{sf}}{dt} \right). \qquad (3)$$

From (3) we obtain the maximum current derivatives

$dI/dt_{max} \sim 5 \div 500$ kA/μs.

Thus, the peak lightning current and the steepness of the lightning current $\Delta i/\Delta t$, which is effective during the interval $\Delta t$, can be related to the final jump phase of the leader discharge. Note that these peak values of the currents and their derivatives are observable in natural Lightning discharges [8-12].

## 4. RADIATION FIELD FROM LIGHTNING LEADER

Results of the measurements of electric and magnetic fields of Lightning contain the information about the distribution of charges and their space-time changing. Therefore determination of the parameters of field source is possible from the measurements of electromagnetic field. It is difficult to solve such inverse problem uniquely. Usually some model of lightning channel is constructed. In the case of simplest model of dipole source the vertical component of radiated electric field has the form of current impulse in a channel:

$$E_{rad}(t) \approx \frac{\mu_0 v_0}{2\pi z} I(t - z/c), \qquad (4)$$

where $\mu_0 = 1.26 \cdot 10^{-6}$ H/m is the magnetic permeability of free space, $v_0$ is the return stroke velocity, $c$ is the light velocity, $z$ is the distance to the lightning strike point.

It is necessary to know functional dependence between different parameters of Lightning in leader stage of discharge development for the construction of adequate model of the return stroke stage. In [6] the following dependence between leader current $I_l$, potential $\varphi_0$, leader velocity $v_l$, streamer zone length $l_{str}$ and return stroke current amplitude $I_0$ were obtained from the condition of stable propagation of a leader:

$$I_l \propto \varphi_l^{3/2} \propto v_l^3 \propto l_{str}^{3/2} \propto I_0 \propto v_0^3 \propto E_{rad}^{3/4}. \qquad (5)$$

In Table 1 corresponding values of these lightning parameters are presented.

Note that the striking distance $r_s$ correlates with the streamer zone length at the final jump phase $r_s \approx l_{str}$. Then it is followed from (2) that $r_s \propto I_0^{2/3}$, that is agreed well with observation data [8].

The electric field derivative $dE/dt$ may be expressed across the parameters of the streamer zone at the final jump phase:

$$\dot{E} \approx \frac{\mu_0 v_0}{2\pi z} \dot{I}(t - z/c) \approx \frac{\mu_0 v_0}{2\pi z} \pi^2 \varepsilon_0 E_s v_{sf}^2. \qquad (6)$$

It is known from the field measurements [13], that the average peak *dE/dt* are about 50 V/m/μs, range-normalized to 100 km, and the corresponding peak current and peak *dI/dt* are about 30÷40 kA and 100 kA/μs. The velocity of the streamer zone front obtained from the expression above is equal to $v_{sf} \approx 40 \div 50$ m/μs, which is consistent with our calculations. Substituting this value for velocity $v_{sf}$ into (2), we obtain the streamer zone length at final jump phase $l_{sf} \approx 20 \div 25$ m.

**Table 1.** Lightning leader parameters and electric field.

| $I_l$, A | $v_l$, cm/μs | $l_{str}$, m | $q_l$, mC/m | $I_0$, kA | $v_0$, m/μs | $E_{rad}$, V/m, z=100 km |
|---|---|---|---|---|---|---|
| 0.91 | 2 | 1.54 | 0.04 | 1 | 24 | 0.05 |
| 24.5 | 6.2 | 13.9 | 0.39 | 27 | 73 | 4 |
| 45.5 | 7.6 | 21 | 0.58 | 50 | 90 | 9 |
| 182 | 12.1 | 52.8 | 1.47 | 200 | 143 | 57 |
| 455 | 16.5 | 97.4 | 2.71 | 500 | 194 | 194 |

Lightning leader parameters are in agreement with the high-speed video observations of cloud-to-ground lightning channels [14-18].

## 5. DISCUSSION AND CONCLUSION

The field derivatives measured in Germany [19] had about 5 time smaller $E_{max}$ values than in the USA [13]. Therefore the re-calculated maximum current steepness differs by a factor of about 5. Such a great discrepancy is often explained by the attenuation effect of the electromagnetic wave propagation along the lossy earth's surface. But such discrepancy takes place also for a distance range from Lightning stroke up to 1 km. This fact cannot be explained by the attenuation effect of the lossy earth's surface. As we showed above the formation of the return stroke current may be explained by the final jump phase of a leader. Because of the lightning discharges in the USA were registered on the see surface then the velocity and streamer zone length at the final jump phase are more greater than in the Germany, where the lightning discharges may be intercepted by the opposite discharges from the highest objects on the earth surface. In this case the velocity and streamer zone length at the final jump phase are substantially smaller.

The final jump phase current depends also on the geometry of the object struck by lightning and whether it is grounded or electrically floating object such as an aircraft [20].

Different electric field waveforms may be observed for positive and negative lightning discharges due to the difference between the leaders propagation mechanism.

Above we considered mainly HF electric fields produced by Lightning. There are also extremely low frequency (ELF) electromagnetic noises produced by Lightning discharge. Recently was shown that these fields may have protective properties for organisms living in stressful conditions [21].

There are two main factors leading to the random oscillations of Lightning electromagnetic field: oscillations of the return stroke current in a Lightning channel and the random bending and branching of breakdown channel [22-24]. Experiments [25] have showed that the oscillations of electromagnetic field of Lightning are caused by the geometry of discharge channel.

Relations obtained between the characteristics of electromagnetic field and parameters of leader and return stroke current allow determination the Lightning parameters from the measurements of the electromagnetic field.

Analysis conducted shows that the formation of the return stroke current front takes place at the final jump phase of a leader and is determined by the length and velocity of the streamer zone of a leader at the final jump.

**References:**

1. M.A. Uman, E.P. Krider, "A review of natural lightning: experimental data and modelling," IEEE Trans., v. EMC – **24**, No. 2, 79-112 (1982).

2. R. Thottappillil, V.A. Rakov, M.A. Uman, "Distribution of charge along the lightning channel: relation to remote electric and magnetic fields and to return-stroke models," J. Geophys. Res. **102**, 6987-7006 (1997).

3. E.N. Chernov, A.V. Lupeiko, N.I. Petrov, "Investigation of spark discharge in long air gaps using Pockels device," Proc. 7$^{th}$ Int. Symp. on High Volt. Eng., Dresden, v.4, 141-144 (1991).

4. N.I. Petrov, V.R. Avansky, N.V. Bombenkova, "Investigation of the characteristics of leader discharge in long air gaps," Proc. 8th Int. Symp. on High Volt. Eng., Yokohama, v.1, 9-12 (1993).

5. N. I. Petrov, V. R. Avanskii, and N. V. Bombenkova, "Measurement of the electric field in the streamer zone and in the sheath of the channel of a leader discharge, " Tech. Phys. **39**, 546 – 551 (1994).

6. N.I. Petrov and R.T. Waters, "Determination of the striking distance of lightning to earthed structures," Proc. Roy. Soc. A. **450**, 589-601 (1995).

7. N.I. Petrov, G.N. Petrova, Modelling of the final jump phase of a positive lightning leader, Proc. Int. Conf. on Gas Discharges and Applications, Cardiff, UK, 509-512 (2008).

8. P.H. Golde, ed. Physics of Lightning, Acad. Press, New York, 1977.

9. E. M. Bazelyan, Yu. P. Raizer, Lightning Physics and Lightning Protection, IOP Publishing, Bristol, Philadelphia, 2000.

10. V.A. Rakov and M.A. Uman, Lightning: Physics and Effects, Cambridge University Press, Cambridge, UK, 2003.

11. M.A. Uman, V.A. Rakov, G.H. Schnetzer, K.J. Rambo, D.E. Crawford, and R.J. Fisher," Time derivative of the electric field 10, 14, and 30 m from triggered lightning strokes," J. Geophys. Res. **105**, 15577-15595 (2000).

12. J.R. Dwyer and M.A. Uman, "The physics of Lightning," Physics Reports **534**, 147-241 (2014).

13. J.C. Willett, E.P. Krider, C. Leteinturier, "Submicrosecond field variations during the onset of first return strokes in cloud-to-ground lightning" Proc. 23th Int. Conf. on Lightning Protection, Firenze, Italy, v.1, p.47-51, 1996.

14. M.M.F. Saba, K.L. Cummins, T.A. Warner, E.P. Krider, et.al., "Positive leader characteristics from high-speed video observation," Geophys. Res. Lett. **35**, L07802 (2008).

15. M. M. F. Saba, L. Z. S. Campos, E. P. Krider, and O. Pinto Jr., "High-speed video observations of positive ground flashes produced by intracloud lightning," Geophys. Res. Lett., **36**, L12811, (2009).

16. L.Z.S. Campos, M.M.F. Saba, T.A. Warner, E.P. Krider, K.L. Cummins, and R.E. Orville, "Does the average downward speed of a lightning leader change as it approaches the ground? – An observational approach," 21st Int. Lightning Detection Conference, 19-20 April, Orlando, Florida, USA, 2010.

17. M. Buguet, P. Lalande, P. Blanchet, S. Pedeboy, P. Barneoud, P. Laroche, "Observation of cloud-to-ground lightning channels with high-speed video camera," XV Int. Conf. Atmos. Electr., 15-20 June 2014, Norman, Oklahoma, USA, 2014.

18. B.M. Hare, O. Scholten, A. Bonardi, S. Buitink, A. Corstanje, U. Ebert, et al. LOFAR lightning imaging: Mapping lightning with nanosecond precision. Journal of Geophysical Research: Atmospheres, 123 (2018).

19. F. Heidler, C. Hopf, "A review about the LEMP characteristics of near return strokes measured in the south of Germany during the last 15 years," Proc. 9$^{th}$ Int. Symp. on High Volt. Eng., Graz, Austria, p. 6742-1-6742-4, 1995.

20. N.I. Petrov, A. Haddad, G.N. Petrova, H. Griffiths and R.T. Waters, "Study of effects of lightning strikes to an aircraft," in Recent Advances in Aircraft Technology, Ed. R.K. Agarwal, Chapter 22, 523-544, 2012.

21. G. Elhalel, C. Price, D. Fixler & A. Shainberg," Cardioprotection from stress conditions by weak magnetic fields in the Schumann Resonance band," Sci. Rep. **9**, 1645 (2019).

22. N.I. Petrov, G.N. Petrova, "Mathematical modeling of the trajectory of a leader discharge and the vulnerability to lightning of isolated and grounded objects," J. Tech. Phys. **40**(5), 427-436 (1995).

23. N. I. Petrov and G. N. Petrova, "Physical mechanisms for intracloud lightning discharges," Tech. Phys. **38**, 287 – 290 (1993).

24. N. I. Petrov and G. N. Petrova, "Physical mechanisms for the development of lightning discharges between a thundercloud and the ionosphere," J. Tech. Phys. **44**, 472 – 475 (1999).

25. D.M. Le Vine, J.C. Willett, "Radiation from Reconstructed triggered Lightning channels," 25$^{th}$ GA URSI, Lille, France, Abstracts, p. 218, 1996.